# Blueprint: Cyberinfrastructure Center of Excellence

**CI CoE Pilot Team**: Ewa Deelman[1], Anirban Mandal[2], Angela P. Murillo[3], Jarek Nabrzyski[5], Valerio Pascucci[4], Robert Ricci[4], Ilya Baldin[2], Susan Sons[3], Laura Christopherson[2], Charles Vardeman[5], Rafael Ferreira da Silva[1], Jane Wyngaard[5], Steve Petruzza[4], Mats Rynge[1], Karan Vahi[1], Wendy R. Whitcup[1], Josh Drake[3], Erik Scott[2]

[1] University of Southern California, Marina del Rey, CA, USA
[2] RENCI, University of North Carolina Chapel Hill, NC, USA
[3] Indiana University, Bloomington, IN, USA
[4] University of Utah, Salt Lake City, UT, USA
[5] University of Notre Dame, Notre Dame, IN, USA

Contact: cicoe-pilot@isi.edu

Version 1.0
Distribution:  Public/Online

---

# Executive Summary

In 2018, the National Science Foundation (NSF) funded an effort to pilot a Cyberinfrastructure Center of Excellence (CI CoE or Center) that would serve the cyberinfrastructure (CI) needs of the NSF Major Facilities (MFs, also known as Large Facilities) and large projects with advanced CI architectures. The goal of the CI CoE Pilot project (Pilot) effort was to develop a model and a blueprint for such a center of excellence by engaging with the major facilities, understanding their CI needs, understanding the contributions the MFs are making to the CI community, and exploring opportunities for building a broader CI community. This document summarizes the results of community engagements that the Pilot conducted during the first two years of the project and describes the identified CI needs of the MFs.

MFs' missions are to deliver data and services to their science and engineering users. The CI supporting these missions is vast and can encompass many components with varied functionality. To better understand that CI, the Pilot has developed and validated a model of the MF data lifecycle (DLC) that follows the data generation and management within a facility and gained an understanding of how this model captures the fundamental stages that the facilities' data passes through from the scientific instruments to the principal investigators and their teams, to the broader collaborations and the public. The Pilot also aimed to understand what CI workforce development challenges the MFs face while designing, constructing, and operating their CI and what solutions they are exploring and adopting within their projects.

Based on the needs of the MFs in the data lifecycle and workforce development areas, this document outlines a blueprint for a Cyberinfrastructure Center of Excellence that will learn about and share the CI solutions designed, developed, and/or adopted by the MFs, provide expertise to the largest NSF projects with advanced and complex CI architectures, and foster a community of CI practitioners and researchers. This blueprint is designed to support the broader NSF vision for CI [1].

This blueprint describes the technical focus for the CI CoE on the MF data lifecycle, the needed expertise, the roles that the Center needs to play in terms of connecting the MFs with each other and the broader CI community, and how that Center needs to develop, adapt, and sustain itself as the CI landscape changes and as the needs of the MFs evolve. The document first describes the assessment methods used to unveil the pressing needs for a CI CoE focused on the MF's data lifecycle; presents the conclusions drawn from these assessments, and elicits the strategies necessary to overcome the identified shortfalls. Note that this blueprint document is not focused on the plan for implementing such a CI CoE, rather it underlines the expected capabilities necessary to serve the MFs' needs.

# 1. NSF Major Facilities

The National Science Foundation and other US agencies have invested significant resources in the development of Major Facilities (MFs) that are critical for US leadership in scientific and economic innovation. The data they acquire and distribute are critical to "harnessing the data revolution" that "plays out over and over again in areas from chemistry to biology to astronomy to physics to engineered systems like Internet of Things (IoT), to education and more" [2]. Although the MFs can differ greatly in their missions, goals, and communities they serve, they all manage and distribute data to advance science [3]. In some cases, the MFs deliver a common set of data products that they collect, analyze, and curate to their community of users and eventually, the public. In other cases, the MFs enable individual principal investigators (PIs) and their teams to collect data using MF's facilities and process it in-house or within the PIs' own environment, and more recently, computing facilities are also being funded as MFs.

> Regardless of what the instruments, data, or processing capabilities within an MF are, cyberinfrastructure is critical to the data collection and its flow from the instruments to the scientists. This CI is complex and changing over time, requiring diverse expertise for its design, construction, and operation. Although the MFs have a significant CI workforce, they usually do not always have the necessary expertise, human resources, or budget to solve every technological issue themselves. At the same time, the knowledge that resides within the MFs is not commonly shared across the MFs or within the broader CI community.

# 2. Assessment Methods

In order to assess the need for a CI CoE focusing on Major Facilities (MFs), we undertook an effort that aimed to achieve the following goals: 1) gain a deep understanding of the CI needs of the NSF Major Facilities, 2) understand the CI needs across the MFs, and 3) understand the role of the planned CI CoE in the context of the broader CI landscape.

Last edited: 12/2/2020During the Pilot phase, the project also defined and initiated three types of engagements, which were structured as follows:

- **Deep engagement**: 1) identify a topic that is important and not-yet fully solved by the MF, 2) conduct preliminary discussions and draft an engagement plan that defines the scope and timeline of activities, sets expectations, and identifies products that will be generated as the result of the engagement, and 3) conduct focused discussions that include a mix of virtual and in-person presence, and hands-on activities such as prototype development.
- **Topical discussions**: 1) identify a specific topic that is important to a number of MFs, 2) facilitate virtual discussions and sessions at conferences, collect and share experiences, distill best practices, and 3) discover opportunities for developing shared capabilities.
- **Community building**: 1) identify related efforts in the broader MF CI community, 2) facilitate and participate in community discussions, organize workshops, 3) collect and disseminate information about the broad community activities, and 4) maintain a living resource for community information.

All the engagements were structured around working groups (WG) [4], with WG leads and members (both from the Pilot effort and MFs). WG were tailored around pressing needs identified in discussions with members of the MFs. WG leads guided focused discussions and provided technical expertise and, when relevant, provided solutions in the form of software proof of concept. WG work products included documents/papers, presentations, videos, prototypes, schema implementations, demonstrations, among others [5].

The Pilot used the vehicle of WG and various engagement modalities to assess the need for a CI CoE. The engagements ranged in time scale (weeks to months) and depth of engagement – from deep engagements with NEON [6], NCAR [7], SAGE [8], and GAGE [9], to topical discussions around identity management, to participation in the Major Facilities and CI for Major Facilities workshops [3], [10], [11]. The Pilot also conducted in-depth interviews with selected MFs, which focused on refining the data lifecycle model and the MF workforce development needs.

4*Blueprint: Cyberinfrastructure Center of Excellence*
*Distribution: Public*

# 3. Assessment of the need for a CI CoE focused on the Major Facility Data Lifecycle

Based on the assessment conducted by the Pilot, the project identified the following areas of need: **1) end-to-end data lifecycle, 2) identity management,** and **3) workforce enhancement and development.** The Pilot saw the need to help MFs to plan for new or improved CI capabilities to support the data lifecycle. The need encompassed technical knowledge as well as operational knowledge: best practices for developing a concept of operations, how to plan for disasters, etc. Some MFs also expressed the need for help in evaluating their capabilities, in some cases in comparison to other technologies and solutions adopted by comparable MFs.

Thus, there is a clear need for an entity to provide the MFs with the expertise to support their end-to-end data lifecycle (DLC), collect and disseminate CI knowledge and solutions related to their DLC, build connections between the CI practitioners within and across the MFs, and build a broader community of CI researchers and practitioners.

## Major Facilities Data Lifecycle

MFs are diverse, highly complex, and heterogeneous. They differ in the types of data they capture, the types of scientific instruments they use, the types of data processing and analysis they conduct, and the policies and methods for data sharing and use. Because of this complexity, the Pilot developed a single lens through which one can reason about the structure and functionality of the CI for MFs and identify specific stages of the data flow within MFs (from data acquisition to dissemination). Through a series of deep and topical engagements with multiple MFs around CI topics, several common data lifecycle stages have emerged. The Pilot then characterized and systematized the MF data lifecycle based on these engagements, additional interviews, ad-hoc discussions, and materials made available to the project. This process has enabled the creation of the facility data lifecycle model [12] that can broadly capture the various stages data goes through a major facility – from scientific data collection to ultimate dissemination and use of the data by scientists. **The data lifecycle model also captures the specific functions and services offered at each DLC stage, the underlying CI supporting each stage, and identifies CI and services that span and impact multiple DLC stages.**

The DLC stages are: 1) data capture/collection, 2) data processing near the instrument(s), 3) data processing at one or more data processing centers, 4) data storage, archiving and curation, 5) data access, dissemination, and visualization. The Pilot also learned that these DLC stages can have unique dependencies among them and often have customized sub-stages, data flows, and loops that are nuanced with respect to particular MFs. In addition, there are data lifecycle aspects that are cross-cutting across these stages, such as: 1) data movement functions, technologies, and policies, 2) data representation, ontologies, and cross-domain data discovery, and 3) operational policies for disaster recovery. Critical to the safekeeping and policy-based sharing of the data are capabilities for identity management for data providers, administrators, and users.

The MFs that the Pilot interacted with during the assessment process have shown the *need for expert advice across the various data lifecycle stages, and at various points in the CI lifecycle* including exploration of solutions, prototyping, system design, operations, and system upgrades and migration. In some cases direct input was needed, in others affirmation of the appropriateness of a chosen solution was helpful.

Since MFs are deeply mission-focused with respect to delivering data to users, the Pilot assessed that they often lack time and resources to do comprehensive research for solutions to CI problems for different DLC stages. Hence, MFs *often look "outward" for expertise and for focused investigation on specific CI topics*, and the *potential application of emerging technology solutions for addressing these MF specific CI problems*. Some of the examples where MFs sought expertise to support their end-to-end DLC have been in the areas of system architecture design for cloud migration of data services, semantic technologies for data management, scientific data visualization, design of sensor pipelines and streaming data platforms, consulting for disaster recovery planning, technologies for identity management, among others.

The Pilot also assessed the *need for transferable CI solutions across MFs for different DLC stages*. Some DLC stages, such as data collection, are unique to each MF, and thus CI solutions in these stages may not be transferable across MFs. However, for many other DLC stages, solutions found for a CI problem within the context of one MF are often applicable to another MF; at the very least the framework for the solutions and the lessons learned in applying the solutions in one MF can be extremely valuable to another MF. Several instances of transferable CI solutions were encountered during the Pilot project, including design of concept of operations, design of messaging systems for data ingest, event-driven data processing, long-term data archiving and curation, data ontology development, etc. During this process, the Pilot identified several potential commonalities and opportunities for amortized leverage (force-multiplier) that a CI CoE can provide to MFs when solving some common technology issues along specific DLC stages and facilitating exchange of CI solutions across major facilities.

The Pilot also assessed the *need for curation of best practices and community knowledge around CI for MF data lifecycle stages*. A majority of MF engagement participants expressed that best practices guidance and existing knowledge base for CI solutions in specific areas of needed expertise will be invaluable at various points in the CI lifecycle, especially during migration, enhancement, or at points where technologies need to be updated. Tutorials, user guides, focused recommendations, reviews and assessments, including the pros and cons of specific CI technologies, have been assessed to be some possible vehicles for curation and dissemination of best practices.

### Identity Management

Identity management (IdM) is a core function that protects all aspects of CI and is embedded in every stage of the MF data lifecycle. Even in the era of open science, the needs in this area are particularly complex, with most MFs authenticating users from multiple institutions [13]. MF's top cybersecurity concerns are "unauthorized or accidental modification of data" and "unauthorized, malicious network/system access" [14], both of which are threats that can be mitigated or exacerbated by identity management that is properly or poorly designed.

Identity management's importance goes beyond cybersecurity implications. NSF and the MFs have a growing interest in understanding who the users of their data are and how they are using that data [14]. Tracing data usage through attractive, user-friendly IdM allows MFs to avoid the ethical concerns of surveillance technologies and the inaccuracies and inconvenience to users of extensive and repetitive surveys.

## Workforce Development Needs

The DLC and its underlying CI within the MFs is very complex and requires quite varied expertise to support the flow of data from remote sensors to data archives, or enable quality data acquisition at sea, or acquire data generated by modern instruments. The MFs need a well trained workforce that is knowledgeable in sensor data management, networking, data management, system management, data archiving, semantic data technologies, web technologies, cloud computing, and many other aspects of CI while also having expertise in the domain served by the MFs. To better understand the workforce needs of the MFs, the Pilot team has set up a number of interviews with MF personnel.

So far, the Pilot team has interviewed fourteen managers from various MFs about workforce development issues. All managers were genuinely interested in seeing their staff develop and grow professionally, but few had outlined a formal approach to assessing training needs, developing employees through mentorship or other professional opportunities, or coaching employees on assessing career goals and methods for realizing those goals. Largely, this was due to the managers having limited support for these activities from the larger organization, but they still had to juggle these tasks while trying to fulfill the technical duties of their individual work, supervise and define direction for their teams, and satisfy competing priorities and needs of their users. For most managers, attending to their staff's professional growth consisted of informally identifying gaps in knowledge that require training, sharing their own knowledge and expertise when able, and encouraging knowledge sharing among the team.

Eight of the fourteen managers discussed opportunities for training offered by their home institution, but many explained that this training was often not advanced enough for their staff's current level of technical expertise or was not even applicable to their duties of managing cyberinfrastructure. Many sought training for their staff elsewhere.

Some managers encouraged cross-training situations to foster redundancy, but this type of training was unavailable, it was most likely not a job-swapping, apprenticeship-style work situation, but more likely an informal transfer of knowledge.

The Pilot also discovered that managers are enthusiastic about opportunities to share knowledge, best practices, and lessons learned with other MFs in regard to workforce development issues, and would appreciate intervention from a CoE in the form of assessments, consultation, offering training, and providing outreach to the broader community of professionals to help clarify misconceptions about MF CI work (which is, admittedly, substantially different than IT work at universities).

# 4. Blueprint for a CI CoE Focused on the Major Facilities Data Lifecycle

Given the needs identified above, there is a demand for a dedicated entity that can serve the MFs in the area of their data lifecycle. In this section, we describe the foremost proficiencies a CI CoE focused on the MF's DLC should provide to assist MFs' decisions regarding CI. Building a CI CoE cannot be done in vacuum, rather the center needs to fit within the broader NSF vision "*for an agile, integrated, robust, trustworthy and sustainable CI ecosystem that drives new thinking and transformative discoveries in all areas of science and engineering (S&E) research and education. The envisioned CI ecosystem integrates advanced CI resources, services and expertise towards collectively enabling new, transformative discoveries across all of science and engineering*" [15]

# CI CoE Focus

A CI CoE needs to primarily serve the requirements of the MFs, re-evaluate them on a periodic basis, and maintain technical and collegial relationships with the members of the MFs. Because there are a number of aspects of CI within the MFs, **the CI CoE needs to focus on a particular facet of the problem, in this case the DLC. To enhance the CI underpinning the DLC, the CI CoE needs to engage with and support the MFs throughout their various phases of the CI lifecycle: planning, construction, operations, and major transitions and upgrades. It can also serve as a partner when the facilities are being transitioned between organizations.**

## Technical Expertise

The CI CoE must have technical expertise in a number of areas associated with the DLC. Examples of expertise areas include, but are not limited to:

- Real time data gathering, filtering, and processing of raw data acquired, for example, from instruments, sensors, etc.; and made available near real-time to the community via data hubs or repositories.
- Distributed, decentralized, or hierarchical models for executing large scale scientific workflow applications and processing large datasets on heterogeneous computing resources and architectures, including high-performance and high-throughput computing systems and clouds; and the ability to provide expert assessment on emerging technologies and architectures.
- Robust and automated data management, visualization, and dissemination, including access, semantics, curation, archiving, and adherence to the FAIR principles.
- Networking (edge-to-cloud) and cloud computing.

In addition to the technology expertise, the Center needs to be able to engage and advise in the areas of CI design, costing, implementation, operations, and upgrades.

The CI CoE must also have a strategy to bring additional expertise as needed to supplement its own capabilities when it is faced with problems it cannot address internally. This strategy should encompass a group of CI experts that are not necessarily active within the Center, but can provide expert knowledge in an ad-hoc manner (e.g., consulting).

Where appropriate, the CI CoE can benefit from partnerships with existing centers of expertise to strengthen the support provided to the MFs. The CI CoE should also have a strategy to Last edited: 12/2/2020update and expand its expertise over time as the CI and the broader technology landscape within academia and industry evolves over time.

## Identity Management

In the area of identity management (IdM), the CI CoE should build on the existing collaboration of the Pilot with Trusted CI [16]. The two projects have collaborated to build a comprehensive set of resources to empower MFs to address their IdM needs, including:
- An IdM Working Group [4], which meets monthly to learn from expert presenters and share experiences, challenges, and lessons learned while implementing IdM at Major Facilities and other NSF research projects.
- One-to-one engagements, in which a joint TrustedCI/CI CoE Pilot team consults with a Major Facility over the course of 3-9 months to solve one or more pressing IdM challenges.
- Outreach [17], training [18], and publications [14] focused on shared problems and solutions being faced at Major Facilities in order to grow institutional knowledge of effective IdM practices across NSF CI operators.
- Resource collection and tools for the NSF CI Community. TrustedCI and CI CoE Pilot maintain a collection of resources and tools [19] for practitioners throughout the community to reference in building their own IdM solutions.

The CI CoE should build on and continue these efforts and periodically jointly survey the community to make sure they are addressing their needs.

## CI Workforce

The CI community and NSF have identified the problem of maintaining and enhancing the CI workforce as one of the critical challenges in supporting science and have launched programs, such as ACI-REF [20] that develops and supports campus facilitators, that assist scientists in using advanced CI. The challenge faced by the MFs is that they not only need to support their communities, but they also need a workforce that can design, build, and operate the advanced CI needed to support their science missions. The CI workforce within MFs is diverse in terms of expertise and backgrounds. Many have degrees in the domain sciences that the MFs serve but they are also required to possess CI expertise. In some cases, the workforce within an MF is fairly static, while the demands of some positions within an MF (like the ship technician) make these jobs relatively short term. Regardless, the needs uncovered in the interviews with fourteen MF managers suggest that a CI CoE needs to offer help to MFs in the form of the following:
- Guide and assist with developing a cohesive strategy for fostering their staff's professional development.

*Blueprint: Cyberinfrastructure Center of Excellence*
*Distribution: Public*

- Help managers understand how to assess development needs and provide coaching to their staff on reaching professional goals.
- Identify and/or provide training opportunities in technical areas as well as "soft skills."
- Guide and assist with developing approaches to cross-training that are perhaps better suited to ensure redundancy than informal transfer of knowledge, and that possibly take advantage of apprenticeship-style opportunities.
- Connect MFs and foster knowledge sharing about workforce development issues.
- Offer services, such as assessments, consultation, training, and outreach/advocacy, with workforce development needs. Specific items mentioned in the interviews included assistance with writing clear and attractive job postings, assessing training, attracting talent with a solid combination of CI and domain science skills, providing a centralized job board for all MFs to share, and providing consultation on managing the complexities of employing non-US citizens.

# CI CoE Roles and Evolution

The CI CoE's role within the broader CI community landscape needs to be well-defined and understood. As time progresses, the outcomes obtained by the Center need to be evaluated and the Center needs to evolve as the CI and the needs of the community change.

## Trusted Partner

Through the discussions with MFs, it became clear that the key to successful communication and collaboration between the MFs and a CI CoE, is trust. Trust has a number of facets including: keeping sensitive information confidential, providing accurate advice based on technical soundness without a particular technological (or other) bias, disclosing conflicts of interest when they exist, conducting interactions in a collegial fashion, and behaving in an advisory capacity, providing constructive comments rather appearing to be auditing or scrutinizing the MFs' activities and solutions.

## Connector

Deep engagements, as implemented by the Pilot effort, are designed to address specific needs of one or two MFs and they last a relatively short time (months). However, there is also a clear need to help connect various MFs with each other [3], which can be accomplished via topical working groups, and with the broader CI community through CI-themed workshops (such as the Major Facilities CI Workshop series) [10]. The various levels of engagement can facilitate knowledge sharing and diffusion across the MFs and across the CI researchers and practitioners.

### Evolution

MFs are long-term NSF investments that will operate over a period of decades. As science advances and CI solutions change, MFs may decide to evaluate and adopt new technologies. The CI CoE can play a critical role in helping the MFs make decisions about their CI, but only if its own knowledge and technical expertise keeps on top of the research and technology trends. To do so, it may also need to train and grow its own workforce from within and also attract new talent to the Center.

### CI CoE as the Community Nexus

As NSF is aiming to fund more CI CoE Pilot and Software Institute efforts, there is an opportunity for the CI CoE to become a nexus for connecting these efforts by: 1) coordinating the organization of joint activities (workshops, training, outreach); 2) increasing the outward visibility of the individual projects; 3) enhancing the CI researcher and practitioner workforce, and 4) building a more connected and stronger community.

As our society changes and as social science clearly demonstrates the benefits of a diverse workforce to the overall health and effectiveness of organizations and our society [21], the CI CoE should play a role in attracting a broad and diverse group of next-generation scientists to the MF and CI workforce.

## Measuring Impact and Planning for the Future

### Outcomes

The CI CoE needs to produce tangible outcomes: engagement reports, publications, presentations, documents, videos, and prototype codes that can serve as a knowledge base for the CoE, the MFs, and the CI community. The distribution of these materials (internal, limited to specific parties, and public) should depend on the content and the intent of the contributors. In some cases, the material may be proprietary for a period of time. Public materials should be made available online. The CI CoE also needs to facilitate discussions between the MFs and the broader CI community by organizing meetings (virtual and in-person) that focus on MF needs. The discussions and outcomes need to be documented and made available to the participants and potentially the public.

## Evaluation

The CI CoE needs to define a set of metrics and criteria under which it will be evaluated by NSF, the MFs, and the CI community. These metrics should be tracked on a continual basis and as a result of an engagement with an MF. The metrics should be a mix of quantitative and qualitative metrics. It may also be beneficial to have an advisory committee that can provide input to the CI CoE on a periodic basis.

## Sustainability

To be effective and to be a true partner for the MFs, the CI CoE needs to be able to sustain itself over time. However, initial NSF funding for seeding the effort and NSF's help in building the relationships with the MFs is a critical component of the CoE. At its inception, the CI CoE Pilot was funded at the same time as NEON was entering its enhancement phase, and it has received funding to launch and sustain this collaboration. It is expected that NSF may have to help initiate additional engagements. However, in the long term, the CI CoE needs to demonstrate value to the MF and CI communities and develop a sustainability plan beyond the initial award.

The sustainability plan should include alignment with the mission and vision of the CI CoE, assessment of strategic partnerships, and evaluation of current sustainability status. Additionally, the sustainability plan needs to include a strategy for capacity building, including the development of community leaders, working group leadership and transitions both in leadership and topics, and continued outreach to new MFs, as well as MFs that are undergoing significant changes. Additionally, the sustainability plan should include a strategy for continued engagements, activities, and services that are adaptable to changes in MFs.

# Acknowledgements

This work was funded by the National Science Foundation under grant #1842042.

The CI CoE Pilot would like to thank our NSF Major Facilities collaborators for participating in the engagements and interviews: **National Ecological Observatory Network (NEON)**: Tom Gulbransen, Dan Allen, David Barlow, Santiago Bonarrigo, Chris Clark, Leslie Goldman, Tristan Goulden, Phil Harvey, David Hulsander, Steven Jacobs, Christine Laney, Ivan Lobo-Padilla, Jeremy Sampson, John Staarmann, Steve Stone; **National Center for Atmospheric Research (NCAR)**: Gordon Bonan, Brian Dobbins, Jim Edwards, David Lawrence, Danica Lombardozzi, Will Wieder, Mariana Vertenstein, **Seismological Facility for the Advancement of**

# References

[1] National Science Foundation, "Transforming Science Through Cyberinfrastructure: NSF's Blueprint for a National Cyberinfrastructure Ecosystem," 2019 [Online]. Available: https://www.nsf.gov/cise/oac/vision/blueprint-2019/nsf-aci-blueprint-v10-508.pdf

[2] "NSF's 10 Big Ideas - Special Report." [Online]. Available: https://www.nsf.gov/news/special_reports/big_ideas/harnessing.jsp. [Accessed: 11-Aug-2020]

[3] E. Deelman *et al.*, "2019 NSF Workshop on Connecting Large Facilities and Cyberinfrastructure: Connecting Large Facilities, Connecting CI, Connecting People," 2019 [Online]. Available: http://dx.doi.org/10.25549/0ZBF-8M77

[4] "CI CoE Pilot: Working Groups." [Online]. Available: https://cicoe-pilot.org/working_groups/. [Accessed: 2020]

[5] "CI CoE Pilot Materials," 2021 [Online]. Available: https://cicoe-pilot.org/materials/

[6] "NSF NEON: Open Data to Understand our Ecosystems." [Online]. Available: https://www.neonscience.org/. [Accessed: 2020]

[7] "NCAR: National Center for Atmospheric Research." [Online]. Available: https://ncar.ucar.edu/. [Accessed: 2020]

[8] "IRIS: Incorporated Research Institutions for Seismology." [Online]. Available: https://www.iris.edu. [Accessed: 2020]

[9] "UNAVCO." [Online]. Available: https://www.unavco.org/. [Accessed: 2020]

[10] "NSF Large Facilities Workshop." [Online]. Available: https://www.largefacilitiesworkshop.com/. [Accessed: 2020]

[11] "2019 NSF Workshop on Connecting Large Facilities and Cyberinfrastructure." [Online]. Available: https://facilitiesci.github.io/2019/. [Accessed: 2019]

[12] L. Christopherson, A. Mandal, E. Scott, and I. Baldin, "Toward a Data Lifecycle Model for NSF Large Facilities," in *Practice and Experience in Advanced Research Computing (PEARC '20)*, 2020, doi: 10.1145/3311790.3396636 [Online]. Available: http://dx.doi.org/10.1145/3311790.3396636

[13] S. Russell, "2019 NSF Community Cybersecurity Benchmarking Survey Report," Trusted CI's Community Survey, 2019 [Online]. Available: https://scholarworks.iu.edu/dspace/handle/2022/24912

[14] R. Kiser, T. Fleury, C. Laney, J. Sampson, and S. Sons, "NEON IdM Experiences," NSF Cyberinfrastructure Center of Excellence (CI CoE) Pilot, 2019 [Online]. Available: https://scholarworks.iu.edu/dspace/handle/2022/23854

[15] "OAC Vision." [Online]. Available: https://www.nsf.gov/cise/oac/vision/blueprint-2019/. [Accessed: 25-Jan-2021]

[16] "Trusted CI: The NSF Cybersecurity Center of Excellence." [Online]. Available: https://www.trustedci.org/. [Accessed: 2020]

[17] "CI/CS Workshop," *Research Security Operations Center (ResearchSOC)*. [Online]. Available: https://researchsoc.iu.edu/training/ci-cs-workshop.html. [Accessed: 2020]

[18] Cyberinfrastructure Center of Excellence-Pilot, "Identity Management Fundamentals," 2020. [Online]. Available: https://www.youtube.com/watch?v=rZWOXeOsN6E

[19] "Trusted CI: Identity and Access Management." [Online]. Available: https://www.trustedci.org/iam. [Accessed: 2020]

[20] "ACI-REF – Advanced CyberInfrastructure – Research and Education Facilitators." [Online]. Available: https://aciref.org/. [Accessed: 11-Aug-2020]

[21] K. D. Gibbs Jr and P. Marsteller, "Broadening Participation in the Life Sciences: Current Landscape and Future Directions," *CBE Life Sci. Educ.*, vol. 15, no. 3, 2016, doi: 10.1187/cbe.16-06-0198. [Online]. Available: http://dx.doi.org/10.1187/cbe.16-06-0198